 \documentclass[final,5p,times,twocolumn]{elsarticle}

\usepackage{amssymb}

\usepackage{graphicx}  %
\usepackage{amsmath}
\usepackage{times}
\usepackage{color}
\usepackage{hyperref}
\usepackage[section]{placeins}
\DeclareMathAlphabet\mathbfcal{OMS}{cmsy}{b}{n}

\journal{Physics Letters B}

\begin{document}

\begin{frontmatter}



\title{Universal physics of bound states of a few charged particles}
\author[1]{C. H. Schmickler}
\ead{christiane.schmickler@riken.jp}
\author[1,2]{H.-W. Hammer}
\ead{hans-werner.hammer@physik.tu-darmstadt.de}
\author[1,3]{A. G. Volosniev}
\ead{artem.volosniev@ist.ac.at}

\address[1]{Institut f\"ur Kernphysik, Technische Universit\"at Darmstadt,
64289\ Darmstadt, Germany}
\address[2]{ExtreMe Matter Institute EMMI, GSI Helmholtzzentrum f\"ur 
  Schwerionenforschung, 64291\ Darmstadt, Germany}
\address[3]{Institute of Science and Technology Austria, Am Campus 1, 3400 Klosterneuburg, Austria}
\date{\today}


\begin{abstract}
  
We study few-body bound states of charged particles subject to
attractive zero-range/short-range plus repulsive Coulomb interparticle forces.
The characteristic length scales of the system at zero energy are set by the Coulomb length scale $D$
and the Coulomb-modified effective range $r_{\mathrm{eff}}$. 
We study shallow bound states of charged
particles with $D\gg r_{\mathrm{eff}}$ and show that these systems
{  obey universal scaling laws different}
from neutral particles.
An accurate description of these states
requires both the Coulomb-modified scattering length and the effective range
unless the Coulomb interaction is very weak ($D\to \infty$).
Our findings are relevant for bound states whose 
spatial extent is significantly larger than the range of the attractive
potential.  These states enjoy
universality -- their character is independent of the shape of the
short-range potential.
\end{abstract}


\end{frontmatter}


\section{Introduction} 
Shallow bound states of two neutral particles with zero angular momentum
live in a classically forbidden region and
retain almost no information about binding interactions~\cite{braaten2006}.
As a consequence, any short-range attractive potential, $V_S$, can be used
to model these states as long as it fixes a few relevant parameters 
(e.g., the scattering length, effective range) to their physical values. 
A celebrated $V_S$ is a zero-range potential tuned to reproduce 
the scattering length~\cite{bethe1935,baz1969,demkov1988}. It provides a powerful starting point
for studying universal bound states (i.e., independent of the shape of $V_S$)
in nuclear and atomic physics~\cite{braaten2006,bethe1935,baz1969,demkov1988,jensen2004,jonson2004,
kraemer2006,tahita2013,kunitski2015,greene2017,naidon2017,hammer2017}. 

In this Letter we consider particles that interact via $V_S+V_C$, where 
$V_C$ is a repulsive Coulomb potential. Potentials $V_S+V_C$ are typical for
cluster models of nuclei~\cite{jensen2004,freer2018,hove2018}, e.g., in $^{17}$F between $^{16}$O and a proton~\cite{morlock1997,Ryberg:2015lea}.
Furthermore, they provide an effective description of interactions between charged quasi-particles, e.g., 
between dressed electrons in crystals~\cite{tulub1958, adamowski1989,Alexandrov1994, devreese2009, kashirina2010,frank2010}.
We focus on $V_S$ of zero range, and explore it
as a possible starting point for understanding realistic charged systems.

The main finding of our study is that there is
  a new family of universal few-body bound states for charged systems
  interacting via a potential $V_C+V_S$. Their properties are
  fully determined  by the Sommerfeld parameter, the Coulomb-modified
  scattering length and effective range, as well as the three-body parameter.
Note that, in contrast to neutral particles, the zero-range approximation to $V_S$ is not guaranteed to be useful
for realistic shallow bound states:
The Coulomb barrier makes the spatial extent of the wave function finite~\cite{fedorov1994}, 
forcing particles to explore the landscape of the short-range binding potential. 
For shallow two-body bound states, we show that 
finite-range corrections to the energy can be accounted for by an effective range parameter.
For weakly-bound three- and four-body systems, we study these corrections numerically using
the Gaussian Expansion Method~\cite{hiyama2003} and Stochastic Variational Method with Gaussians~\cite{varga1995,suzuki1998,suzuki2002,mitroy2013}.
{  Details of the numerical methods are given in \ref{app:num}.}

\section{Two-Body System}
We consider two particles whose relative motion is described by the radial Schr{\"o}dinger equation
\begin{equation}
  -\frac{\hbar^2}{2\mu}\frac{\partial^2 u}{\partial r^2}
  +\left[V_{C}(r)+V_{S}(r)\right]u(r)=-\frac{\hbar^2 \kappa^2}{2\mu} u(r),
\label{eq:schr_two}
\end{equation}
where $\mu$ is the reduced mass, $E_2=-\frac{\hbar^2 \kappa^2}{2\mu}$ with $\kappa>0$ is the two-body energy, 
$V_{C}=k_e\frac{Q_1Q_2}{r}$ is the Coulomb potential energy ($Q_1$, $Q_2$ are the particle charges, $k_e$ is Coulomb's constant), and $V_S$ is a binding potential of range $R$. 
We consider only zero angular momenta since our focus is on the bound states for $R\to 0$ (later 
referred to as the zero-range or universal limit).
Moreover, we are free to choose any shape of $V_S$, which is irrelevant as long as the limit $R\to 0$ is well-defined 
for neutral particles interacting via $V_S$ (cf.~\cite{baz1969, landau1977}).
For simplicity, we assume that $V_C+V_S$ is a square well for $r\leq R$, i.e.,
$V_S(r) =-\frac{\hbar^2 g}{2\mu R^2}-V_C(r)$, and $V_C$ otherwise. 
The dimensionless parameter $g>0$ sets the interaction strength. 
 The wave function $u$ for this potential reads
\begin{equation}
 u(r) =
 \cal{N}\times \begin{cases}
    \sin\left(\frac{r}{R} \sqrt{g-\kappa^2R^2}\right)     &  \text{if } r\leq R\\
    W_{-\eta,1/2}(2\kappa r)\frac{\sin\left(\sqrt{g-\kappa^2R^2}\right)}{W_{-\eta,1/2}(2\kappa R)} &  \text{if } r>R,
  \end{cases}
\label{eq:wave_func}
\end{equation}
where $\cal{N}$ is a normalization constant that ensures $\int_0^\infty u^2\mathrm{d}r=1$, $\eta\equiv \frac{k_e\mu Q_1 Q_2}{\hbar^2 \kappa}$ is the Sommerfeld parameter,
and $W$ is the Whittaker $W$-function~\cite{abramowitz1964}.
The values of $\kappa$ that lead to a continuous derivative of $u$ at $r=R$ define
allowed bound states.
In the limit $R\to 0$, $\kappa$ is a root of the equation
\begin{equation}
\sqrt{g}\cot(\sqrt{g})=(2\eta\psi(\eta)+2\eta \ln(2\kappa R)+4\eta\gamma+1)\kappa R,
\label{eq:non_universal_energy}
\end{equation}
where $\gamma$ is Euler's constant, and $\psi$ is the digamma function~\cite{abramowitz1964}.
Note that for neutral particles ($\eta=0$) 
Eq.~(\ref{eq:wave_func}) depends only on $\kappa R$, hence, the result of taking 
the limit $R\to 0$ with fixed $\kappa$ is identical to that with $\kappa \to 0$ and fixed $R$. 
In other words, for weakly-bound states of neutral particles one may always rely on the zero-range limit. 
For charged particles with $\kappa\to 0$, by contrast, the zero-range limit is not necessarily accurate. 

{  We rewrite
  Eq.~(\ref{eq:non_universal_energy}) by following Ref.~\cite{landau1977}. 
Collecting the energy-independent terms,
we require their sum to approach the Coulomb modified scattering length,
\begin{equation}
  a_C=D\left[2 \ln\left(\frac{2R}{D}e^{2\gamma}\right)-\sqrt{g}\cot(\sqrt{g})\frac{D}{R}\right]^{-1}\,,
\end{equation}
where
\begin{equation}
  D=\frac{1}{\kappa\eta}=\frac{\hbar^2}{k_e\mu Q_1 Q_2}
  \end{equation}
is the length scale associated with the Coulomb potential
(i.e., the generalized Bohr radius).
This leads to a model-independent relation for $R\to 0$,
\begin{equation}
2\eta\psi(\eta)-2\eta \ln(\eta)+1=-\frac{1}{\kappa a_C},
\label{eq:universal_energy}
\end{equation}
which connects one observable (the binding energy) 
to another (the scattering length).}
Previously, Eq.~(\ref{eq:universal_energy}) was derived
for proton-proton 
states~\cite{landau1944,jackson1950,Kong:1998sx,Kong:1999sf}. It is related to
the poles of the scattering amplitude defined as in Refs.~\cite{holstein1999,higa2008}. 

\begin{figure}
\begin{center}
\includegraphics[scale=0.3]{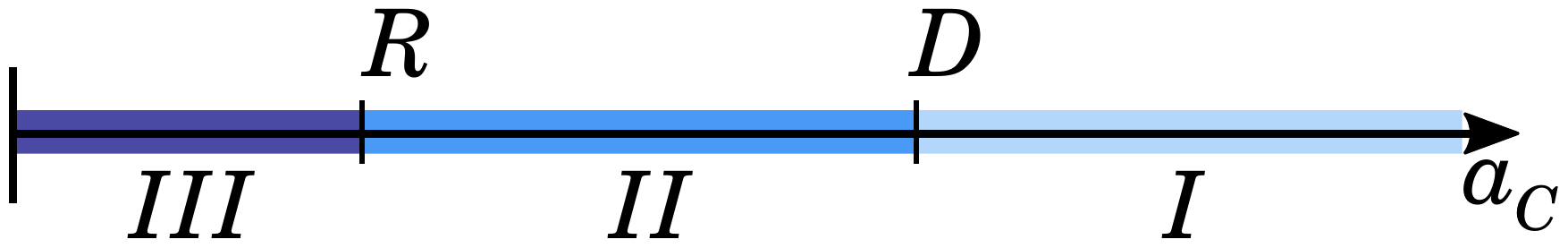}
\end{center}
\caption{Different ranges of the Coulomb-modified scattering length
  $a_C$ in relation to the range, $R$, of $V_S$ and
  the Coulomb length, $D$.}
  \label{fig:fig1}
\end{figure}
In this Letter, we consider Eq.~(\ref{eq:universal_energy}) 
  in the context of universal bound states which require $D \gg R$.
  We end up with the three different regions in Fig.~\ref{fig:fig1}
  for the location of the scattering length
  $a_C$ with respect to the range, $R$, of $V_S$ and
  the Coulomb length, $D$. We consider the regions I ($a_C \gg D$) and
  II ($R \ll a_C \ll D$) where universal physics can be expected.
  (In region III this is not likely because the system probes the shape of $V_S$.)
The left-hand-side of Eq.~(\ref{eq:universal_energy}) 
is a monotonic negative function. Therefore,
if $a_C>0$ a zero-range potential can support 
at most one bound state\footnote{ This statement is correct only for zero-range potentials. 
A finite range potential can support any finite number of bound states. Depending on the prescrition
for taking the limit $R\to 0$, those states either vanish or become infinitely deep.}, and if $a_C<0$ there can be no bound
states.
In region II (``weak Coulomb''), we obtain
\begin{equation}
\kappa\simeq \frac{1}{a_C}\left[1-\frac{2 a_C}{D}\ln\left(\frac{a_C e^{\gamma}}{D}\right)\right].
\label{eq:eta_0}
\end{equation}
This equation features a logarithmic correction to the standard
expression $\kappa=1/a_C$ for neutral particles~\cite{braaten2006}. { Logarithmic dependence 
on $D$ is typical for ``weak Coulomb'~\cite{jackson1950, Barford:2002je}.}
In region I (``strong Coulomb'') we have
\begin{equation}
\kappa^2\simeq \frac{6}{D a_C}+\frac{18}{5 a_C^2},
\label{eq:eta_large}
\end{equation}
which describes shallow bound states. The fact that 
$\kappa^2\sim 1/a_C$ in the limit $a_C\to\infty$ will be of utmost 
importance for finite-range corrections. We now focus on this new class
of shallow states.

To calculate other observables, we note that for $R\to 0$ 
the particles move almost exclusively in the classically forbidden region. 
Indeed, the probability to find particles with $r>R$ is approaching unity:
$P(r>R)=1-\int_0^R u^2 \mathrm{d}r\xrightarrow[]{R\to 0} 1\,$.
To derive this limiting value, we notice that 
${\int_0^R u^2 \mathrm{d}r} < {u^2(R)R}$ for $R\to 0$.
Therefore, observables for zero-range interactions are described with the wave function $W_{-\eta,1/2}(2\kappa r)$,
defined by $\eta$ and $\kappa$.
As an example, we use the root-mean-square (rms) radius,
$\langle r^2 \rangle\equiv \int u^2 r^2\mathrm{d}r$, -- a standard observable
in few-body physics -- given by 
\begin{equation}
\frac{\sqrt{\langle r^2 \rangle_0}}{D} = \eta\sqrt{\frac{\int_{0}^\infty W^2_{-\eta,1/2}(2 x) x^2 \mathrm{d}x}{\int_{0}^\infty W^2_{-\eta,1/2}(2 x) \mathrm{d}x}},
\label{eq:r2_u}
\end{equation}
where the subscript $0$ refers to the zero-range limit. The right-hand-side of Eq~(\ref{eq:r2_u})
is a monotonically increasing function of $\eta$. The maximum value is attained
at $1/\eta = 0$ where $\sqrt{\langle r^2\rangle_0}/D=0.507\dots $;
in this limit the size of the bound state is fully determined by $D$.
The boundedness of the rms radius is relevant for charged halo nuclei~\cite{fedorov1994}; it also
supports the predicted discontinuous behavior of the mean distance between 
two polarons across the unbound-polarons to bipolaron transition~\cite{adamowski1989, verbist1991}.
For $\eta\to 0$ the rms radius  is determined from the standard relation
 $2 \langle r^2 \rangle_0 \kappa^2\to 1 $~\cite{jensen2004}.

\section{Finite-range corrections}
The rms radius  in Eq.~(\ref{eq:r2_u}) does not diverge for $\kappa\to 0$, indicating
that the inclusion of finite-range corrections is unavoidable for charged systems. 
These corrections must be small
if $\sqrt{\langle r^2 \rangle} \gg R$, 
which, according to Eq.~(\ref{eq:r2_u}), is satisfied for weakly-bound systems if $D\gg R$. 
For comparison, $D\simeq 57.6$~fm for two free protons and $D \simeq 0.1$~nm for free electrons.
Therefore, if two protons (or a proton with a light nucleus) formed a shallow bound state 
it would be universal,\footnote{ One might speculate that proton-proton correlations
in a nuclear medium (e.g., in the outer core of a neutron star) 
can potentially lead to states relevant for our results.
Since these correlations are not fully understood~\cite{baldo2007},
we omit this discussion here.}
since natural values of $R$ in this case are around $1$~fm. Two dressed electrons in solids could represent 
another universal system where the effective mass, the strength of the Coulomb potential (hence $D$),
as well as $R$ depend on the material. In contrast, a shallow bound state of two free atomic ions
($D\simeq 57.6$~fm and $R\sim 0.1$~nm) cannot be universal. 

\begin{figure}[tb]
\includegraphics[scale=1]{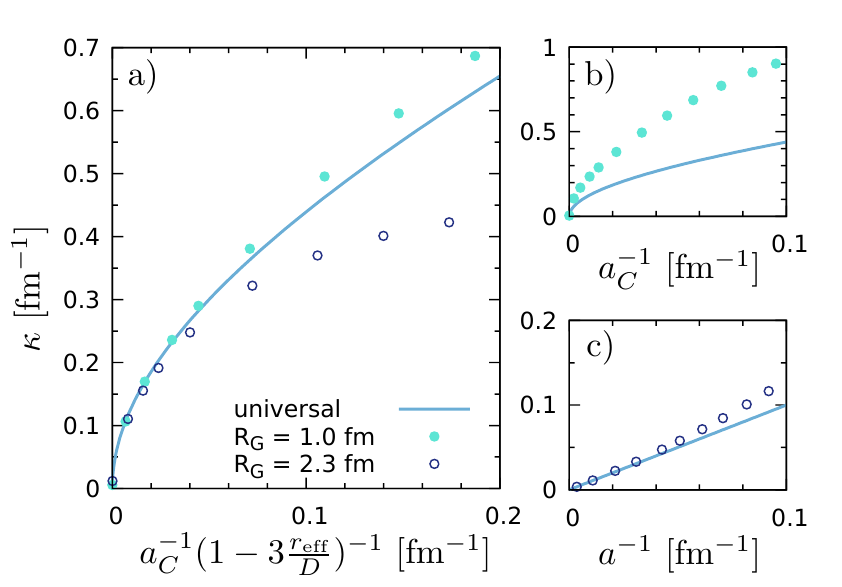}
\caption{Panel {\bf a)}: $\kappa$ as a function of the inverse of the rescaled Coulomb-modified scattering length $[a_C(1-3r_\mathrm{eff}/D)]^{-1}$.
The solid curve shows the universal limit for charged particles.
The circles present results for the Gaussian potential of the range $R_{G}$ (see the legend).
For the sake of discussion, we use the masses and charges of two alpha particles. 
Panel {\bf b)}: $\kappa$ as a function of $1/a_C$. The notation is as in {\bf a)}. 
Panel {\bf c)}: For comparison, we plot $\kappa$ as a function of the inverse scattering length $1/a$ for neutral particles (all other parameters are as in {\bf a)}).}
\label{fig:energy_two_body}
\end{figure}

To estimate finite-range corrections to the energy, we notice that 
the right-hand-side of Eq.~(\ref{eq:universal_energy}) is the first term 
of the Coulomb-modified effective-range expansion. To account for the next term,
one must use $-1/(a_C\kappa)-r_{\mathrm{eff}}\kappa/2$ instead of $-1/(a_C\kappa)$,
where $r_{\mathrm{eff}}$ is the effective range~\cite{bethe1949,goldberger1964}
\begin{equation}
  2\eta\psi(\eta)-2\eta \ln(\eta)+1=-\frac{1}{\kappa}\left( \frac{1}{a_C}
  +\frac{r_{\mathrm{eff}} \kappa^2}{2}\right).
\label{eq:energy_correction}
\end{equation}
The effective-range correction enters at
leading order in the expansion of the energy for weakly bound states
\begin{equation}
  \kappa^2\simeq \frac{6}{a_C D (1- 3 r_{\mathrm{eff}}/D)}\,,
  \label{eq:kappaq}
\end{equation}
because the leading contribution to $\kappa^2$ in
Eq.~(\ref{eq:eta_large}) is proportional to $1/a_C$.
{ We note that Eq.~(\ref{eq:kappaq}) was previously
  derived in the context of connecting
  asymptotic normalization constants of charged bound states to scattering
  parameters~\cite{Sparenberg:2009rv}.}
Higher order finite-range
corrections (e.g., due to the shape parameter)
are not as important in the limit $1/a_C\to 0$ because they are convoluted with 
$\kappa^n$ where $n>2$.
Note that the factor $(1- 3 r_{\mathrm{eff}}/D)$ in the denominator of $\kappa^2$
for $1/a_C\to 0$
implies that for potentials with weakly-bound states
$r_{{\mathrm{eff}}}<D/3$ must hold, in agreement with the causality
constraints of~\cite{koenig2013}.

Our result can be used to define the leading order of an
effective field theory for shallow bound states of charged particles
where $r_{\mathrm{eff}}$ contributes at leading order, while higher
effective range parameters can be included perturbatively. The fact
that range corrections are enhanced in systems with strong Coulomb
interactions was already observed for $^{17}$F~\cite{Ryberg:2015lea}
and $^7$Be~\cite{Higa:2016igc}, and attributed to an additional
fine tuning. The importance of finite-range effects in systems with
strong Coulomb interactions was also observed
in~\cite{papenbrocktalk,Luna:2019ufu}.
Here, we show that this enhancement is generic for ``strong Coulomb''.
Effective field theories designed for bound systems in region II that
  use only $a_C$ as two-body input in the leading order
  (see, e.g.,~\cite{birse2010, ryberg2014}) must be extended to describe
  shallow-bound nuclei close to the proton dripline where 
$r_{\mathrm{eff}}/D$ is not  small.

To illustrate finite-range effects, we use the Gaussian potential
{   \begin{equation}
    V_S^{G}=g_G e^{-r^2/(2 R_G^2)}\,,
  \end{equation}
}
where $R_G$ defines the range of the potential,
and $g_G$ is used to fix the Coulomb-modified scattering length for 
a fixed value of $R_G$.  
For the sake of discussion, we use parameters of two 
$\alpha$-particles $\hbar^2/\mu=20.73$ MeV$\times$fm$^2$ and $kQ_1 Q_2=5.76$ MeV$\times$fm [$D\simeq 3.6$~fm].
We employ the Gaussian Expansion Method~\cite{hiyama2003}
to calculate $\kappa$. The result is plotted in Fig.~\ref{fig:energy_two_body}{\bf a)}
as a function of $[a_C(1-3r_{\mathrm{eff}}/D)]^{-1}$ (for consistency, we take $r_{\mathrm{eff}}$
for $V_S^G$ with $\kappa=0$), which is the only relevant 
parameter for $1/a_C\to\infty$; see Eq.~(\ref{eq:energy_correction}).
Figure~\ref{fig:energy_two_body} shows that even though the universal prediction 
does not describe finite values of $r_{\mathrm{eff}}/D$ (see Fig.~\ref{fig:energy_two_body}{\bf b)}), 
it is still useful:
Finite-range corrections for shallow bound states are captured by
rescaling $a_C$ with $1-3r_{\mathrm{eff}}/D$, see
Fig.~\ref{fig:energy_two_body}{\bf a)}.

Expectation values of other observables $(\langle O\rangle \equiv \int u^2 O \mathrm{d}r)$
also acquire finite-range corrections when $r_{\mathrm{eff}}\neq 0$.
It is particularly easy to calculate these corrections for an observable $O$, which in the limit $R\to 0$
satisfies $\int_{0}^R O u^2 \mathrm{d}r \sim R^{1+\delta}$, where $\delta>0$, e.g., the rms radius. 
After straightforward but tedious calculations we derive (in the leading order in $r_{\mathrm{eff}}/D$)
\begin{equation}
\frac{\langle O\rangle}{\langle O\rangle_0}\simeq 
1+\frac{r_{\mathrm{eff}}}{2D}\frac{W^2_{-\eta,1/2}(0)}{\int_{0}^\infty W^2_{-\eta,1/2}(2x)\mathrm{d}x},
\label{eq:Observables}
\end{equation}
where $\langle O\rangle_0$ is the universal prediction for the same $\kappa$ and $\eta$.
As anticipated, the universal value $\langle O\rangle_0$ is accurate if $R \ll D$ 
(note that $r_{\rm eff}\sim R$ for $D/R \gg 1$). 
The left-hand-side depends weakly on the energy ($\eta$) and for $\eta > 2$ 
it can be accurately written as $1+\frac{r_{\mathrm{eff}}}{2D}(6+\frac{1.1}{\eta^2})$.
Note also that the correction in Eq.~(\ref{eq:Observables}) is independent of $O$. Therefore, 
it can, in principle, be used to relate different measurements.

{ Finally, we discuss what happens to a shallow bound state if there is a small perturbation to $V_S$ that changes the sign of the scattering length, i.e., $1/a_C= 0^-$ in a new potential. Such a perturbation modifies the nature of a discrete state. The bound state turns into a resonance whose width is determined by the probability for a particle to overcome the Coulomb barrier. This probability is given by the Gamow-Sommerfeld factor, $P_g=\exp[-2\pi/(Dk)]$, where $k>0$ is the scattering wave vector, ($\kappa=i k$ in Eq. (1)). 
To make the discussion more quantitative, we expand the Coulomb-modified phase shift~\cite{landau1977} for $k\to 0$
\begin{equation}
\cot(\delta_0)=-\frac{1}{2\pi P_g}\left[\frac{D}{a_C}+\frac{k^2D^2}{6}\left(1-3\frac{r_{\mathrm{eff}}}{D}\right)\right] +... \;.
\end{equation}
If $a_C<0$ and large, then low-energy scattering shows a resonance feature that extends Eq.~(\ref{eq:kappaq}) to $a_C<0$.
If $a_C>0$ or $a_C<0$ and small, then $\delta_0\sim P_g$, which is exponentially small for $k\to 0$.
In this case, the scattering states are decoupled from the short-range potential;
all scattering observables are accurately determined by $V_C$.}

\section{Three-Body System, Efimov effect, and Thomas collapse}
Now we consider a three-body system of charged particles with
mass $m_i$, charge $Q_i$, and coordinates $\mathbf{r}_i$, $i=1,2,3$,
interacting via Coulomb and short-range pair interactions as in
Eq.~(\ref{eq:schr_two}).
For $Q_i \equiv 0$, this system features two hallmarks of few-body physics: 
the Efimov effect~\cite{efimov1971} and the Thomas collapse~\cite{thomas1935}.
Both are connected in the hyperspherical formalism~\cite{nielsen2001} to 
a super-attractive $\rho^{-2}$-potential in the hyperradius,
$\rho^2=(r_{12}^2+r_{13}^2+r_{23}^2)/3$, with $r_{ij}=|\mathbf{r}_i-\mathbf{r}_j|$.
The Thomas collapse 
occurs due to the divergence of $1/\rho^2$ at the origin~\cite{landau1977}, 
whereas the infinite tower of Efimov states is supported by the scale invariance of this potential.
The  $1/\rho^2$ form of the potential strongly suggests that the
Thomas effect is weakly modified by the Coulomb potential, but the shallow
Efimov states must disappear.  All in all, this indicates
that universal systems of charged particles with zero-range interactions do
{  obey universal scaling laws \cite{jensen2004,braaten2006}
different from}
neutral particles.

\begin{figure}
\includegraphics[scale=1]{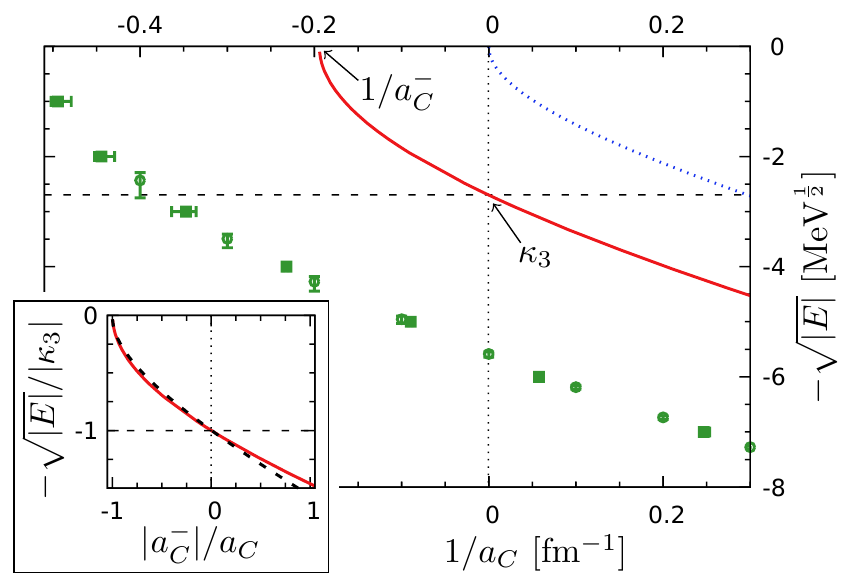}
\caption{Energies of few-body states of charged bosons in the zero-range limit.
  {  Mass and charge of the bosons are taken from alpha particles
 [$D\simeq 3.6$~fm].}
The dotted curve shows the two-body result
of Eq.~(\ref{eq:universal_energy}).
The solid curve is the three-body energy. The dots with error bars present the
extrapolation to the universal four-body results. 
The three-body energy (determined by $\kappa_3$) at $1/a_C = 0$ is set by the three-body force.  
The value $a_C^-$ below which the three-body state does not exist is also shown.
The inset shows the rescaled three-body energies for different values of the three-body force.
}
\label{fig:fig3}
\end{figure}

{  In order to elucidate the nature of the universal scaling laws
  in the presence of Coulomb interactions, we investigate the  fate of the
  Efimov effect and Thomas collapse which are hallmarks of universality
  for neutral systems. 
We first show the absence of the Efimov effect in a system of two identical heavy charged particles interacting with a light neutral particle, i.e., 
$m_1=m_2, m_3/m_2\ll 1, Q_1=Q_2$ and $Q_3=0$.
This system is conveniently studied within the Born-Oppenheimer approximation: 
First, the Schr{\"o}dinger equation with heavy particles fixed at $\mathbf{r_1}$ 
and $\mathbf{r_2}$ is solved,
\begin{equation}
\left[-\frac{\hbar^2}{2m_3}\frac{\partial^2}{\partial {\mathbf{r_3}}^2}
+V^{lh}_S(\mathbf{r}_1-\mathbf{r}_3)+V^{lh}_S(\mathbf{r}_2-\mathbf{r}_3)\right]\Psi^{lh}=\epsilon \Psi^{lh},
\end{equation}
which gives the energy $\epsilon(|\mathbf{r}_1-\mathbf{r}_2|)$.
The superscript $lh$ is used to emphasize that the equation describes the light-heavy subsystem.
Then, the energy spectrum is found from the two-body equation
\begin{equation}
\left[-\frac{\hbar^2}{m_1}\frac{\partial^2}{\partial {\mathbfcal{R}}^2}+
\epsilon(\mathbfcal{R})+V_C(\mathbfcal{R})+V^{hh}_S(\mathbfcal{R})\right]\Phi^{hh}=E\Phi^{hh},
\end{equation}
where $\mathbfcal{R}\equiv \mathbf{r}_1-\mathbf{r}_2$, and $\Phi^{hh}$ is the wave function that describes 
the relative motion of the two heavy particles. We are interested in the behavior 
of the potential, $\epsilon+V_C+V^{hh}_S$, at $\mathbfcal{R}\to \infty$. Therefore, without loss of generality, we assume that (i) $V^{hh}_{S}$ is an infinite
barrier for $|\mathbfcal{R}|<R$ and zero otherwise; (ii) $V^{lh}_{S}$ is a separable s-wave interaction.
The assumption (ii) allows us to write the function $\epsilon(\mathbfcal{R})$ analytically~\cite{fonseca1979}.
}
For an infinite heavy-light scattering length this function reads: 
$\epsilon(\mathbfcal{R}\to\infty)\simeq-\frac{\hbar^2 A}{m_1 \mathbfcal{R}^2}$ 
where $A>1/4$ generates infinitely many bound states if $V_C=0$.
For charged particles the Schr{\"o}dinger equation,
\begin{equation}
\left(-\frac{\hbar^2}{m_1}\frac{\partial^2}{\partial \mathbfcal{R}^2}
-\frac{\hbar^2 A}{m_1 \mathbfcal{R}^2}+k\frac{Q_1^2}{|\mathbfcal{R}|}\right)\Phi=E\Phi, 
\label{eq:mod_Efimov}
\end{equation}
cannot support infinitely many-bound states. It can support at most $N$ bound states. $N$ 
can be estimated using the Bargmann inequality~\cite{bargmann1952}:
\begin{equation}
  N\leq \frac{2(R-b)}{D}+A\ln\left(\frac{A D}{2 R}\right)\,,
\end{equation}
where $D$ is the Coulomb length for two heavy particles.
It is clear that only if $D/R\gg 1$ there can be many bound states.  
For example, for the parameters as in $^9$Be described as an $\alpha+\alpha+n$
system~\cite{hiura1972,fonseca1988}
we have $N\lesssim 1.5$, where, for simplicity, we used natural values:
$A=1.25$ and $R=1$~fm. 

{ We now move on to charged particles with equal masses.
It can be shown that the Coulomb potential also dominates the  long-range behavior 
of the lowest adiabatic potential in the hyperspherical formalism (cf.~\cite{fedorov1996}), which leaves 
no room for the Efimov effect with identical charged bosons. However, the
low-lying Efimov states survive if the Coulomb interaction is sufficiently weak~\cite{schmickler2019}.
Our interest is in these states. 

We study the Thomas collapse numerically (see \ref{app:thomas} for more
details) and observe that the ground state energy
behaves similarly to that for neutral particles, i.e., $E\sim -1/R^2$, 
in the vicinity of the zero-range limit ($R\to 0$).
This is consistent with the findings of Ref.~\cite{birse2010}
   in their non-perturbative treatment of $^3$He.
Therefore, to study three-body states in the universal limit, 
we introduce a three-body potential
\begin{equation}
V_{{\mathrm{3b}}}=g_{{\mathrm{3b}}}e^{-(r_{12}^2+r_{23}^2+r_{13}^2)/(16 R_G^2)}
\nonumber
\end{equation}
where $g_{{\mathrm{3b}}}$ is chosen to fix the three-body energy.}
As we show below, this three-body force also allows us to study 
a four-body problem without any additional parameters. 
This is similar to neutral systems~\cite{platter2004,yamashita2006,stecher2009},

\section{Universal three- and four-body states} 
We use the Gaussian Expansion Method~\cite{hiyama2003} to study few-body states in the zero-range limit.
We obtain the energies by performing a sequence of calculations with small values of $R_G$ 
and extrapolate to $R_G\to 0$, which gives the value at $R_G=0$ and the error bars. {  Details of the extrapolation procedure are given in
  \ref{app:extra}.} 
Figure~\ref{fig:fig3} reports on the energies of two, three and four charged bosons.
As before, the mass and charge of a boson are those of an alpha particle. 
The energies are fully determined by $D$, $a_C$ and an additional three-body parameter. 
The latter can be characterized either by $a_C^-$ which determines the three-body binding threshold or by the three-body energy 
at $1/a_C=0$, $\kappa_3$ (see Fig.~\ref{fig:fig3}).
The energy of the three-body state at $1/a_C=0$, and, hence, $\kappa_3$, is fixed by the three-body force. 
For neutral particles another value of $\kappa_3$ would simply rescale the $y$-axis and $x$-axis
due to the discrete scale invariance. For charged particles the discrete scale invariance is broken (cf. Eq.~(\ref{eq:mod_Efimov})).
Thus, we also should investigate the effect of the three-body force; see the inset of Fig.~\ref{fig:fig3}. 
We use two different three-body forces whose $a_C^-$ differ by a factor of 10, and then rescale the $x$-axis 
and $y$-axis using $a_C^-$ and $\kappa_3$, correspondingly. 
We see that
the effect of the three-body parameter leads to merely a rescaling 
of the axes for the considered cases. Therefore, we refrain from showing energies for other values of $a_C^-$. 

\begin{figure}
\includegraphics[scale=1]{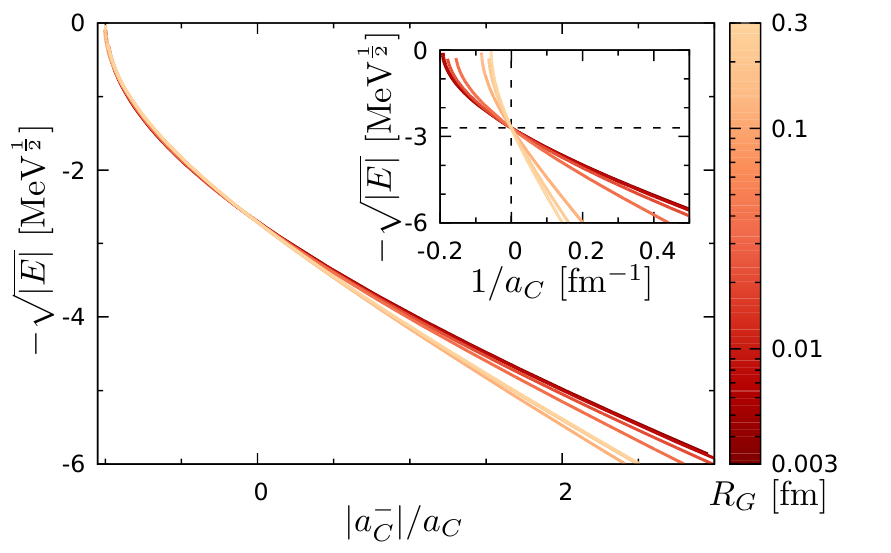}
\caption{Energies of the trimer for different values of $R_G$ as functions of $|a_C^-(R_G)|/a_C$. { The mass and charge are taken from alpha particles
[$D\simeq 3.6$~fm]. The darker curves correspond to smaller values of $R_G$, while
the three-body force is chosen such that all curves intersect at $1/a_C=0$.
The inset shows the energies as functions of $1/a_C$.}}
\label{fig:figure4}
\end{figure}

We study finite-range corrections numerically; see Fig.~\ref{fig:figure4}. 
{  Even values of $R_G\ll D$ immediately lead to 
significant corrections to the energy;} see the inset of Fig.~\ref{fig:figure4}. 
Similarly to two particles, these corrections can be accounted for by rescaling 
$a_C$. To demonstrate this, we define $a_C^-(R_G)$ that determines the three-body 
binding threshold for a given value of $R_G$. We use $a_C^-(R_G)$ to rescale the $x$-axis.
The rescaled curves coincide for $a_C<0$, which suggests that the universal limit 
is a good starting point for studying Borromean three-body charged systems. 

\section{Outlook}
We have demonstrated that the universal physics of shallow bound states of charged
particles is different from neutral particles.
The characteristic scale of the system is set by the Coulomb length scale $D$.
Unless the Coulomb interaction is extremely weak ($D\to\infty$), the effective range
is needed to determine the universal properties of shallow bound states.
More detailed studies are required to see in which systems the features  
discussed here can be observed.  
To this end, charged quasi-particles and nuclei close to the proton 
drip-line~\cite{kashirina2010,woods1997,adamowski1989,lewis1999,hagen2010,
morlock1997,Ryberg:2015lea,fedorov1994,geithner2008,fedotov2010,parfenova2018}
must be investigated. 
The concept of universality for neutral particles has also been
explored in low and mixed spatial
dimensions~\cite{bruch1979,hammer2004,armstrong2010, volosniev2011,tan2012,nishida2018},
and with higher angular momenta~\cite{kartavtsev2007,nishida2013,volosniev2014}.
It will be interesting to study the effect of the Coulomb potential 
on those universal states, especially in connection with 
low-dimensional bipolarons~\cite{takada1982, emin1992,vansant1994},
 p-wave halo nuclei such as $^8$B~\cite{schwab1995,Zhang:2015ajn},  and exotic 
states of $\alpha$ particles~\cite{fynbo2011}.
Finally, it would be interesting to formulate an
effective field theory for shallow bound states of charged particles
based on our findings and calculate corrections from higher
effective range parameters perturbatively.

\vspace*{1em}

We thank Wael Elkamhawy for comments on the manuscript, and Daniel Phillips
for pointing us to~\cite{papenbrocktalk}.
  We also thank Michael Birse, Wael Elkamhawy, Dmitri Fedorov, Aksel Jensen, Daniel Phillips, and Karsten Riisager for useful discussions
 about proton halos, and Emiko Hiyama for many inspiring conversations about the Gaussian Expansion Method. 
  This work has been supported by the Deutsche Forschungsgemeinschaft
  (DFG, German Research Foundation) under project numbers 
  413495248 -- VO 2437/1-1 and 279384907 -- SFB 1245 and by the Bundesministerium
  f\"ur Bildung und Forschung (BMBF) through contract 05P18RDFN1.

\appendix

\section{Numerical Methods}
\label{app:num}

We employed the Gaussian expansion method~\cite{hiyama2003} for numerical calculations in the main text.
It is a variational method in which the wave function is expanded as a sum of Gaussians. In this section we briefly 
illustrate the method for a three-body system; see~\cite{schmickler2017, schmickler2019}
for a more detailed presentation.
A variational wave function is written as 
\begin{equation}
\Psi_{GEM}=\sum_{i=1}^M a_i \mathcal{S} e^{-\alpha_i x^2-\beta_i y^2},
\end{equation}
where ${\bf x}={\bf r}_1-{\bf r}_2$ and ${\bf y}={\bf r}_{3}-({\bf r}_1+{\bf r}_2)/2$ are the Jacobi coordinates,
$\mathcal{S}$ is a symmetrization operator [the main text considers only spinless bosons],
and $M$ is the basis size.
The parameters $\alpha_i$ and $\beta_i$ are chosen in a form of a geometric progression, i.e., $
\alpha_i=\alpha_1 A^{i-1}$ and $\beta_i=\beta_1 B^{i-1}$, where $A$ and $B$ are input parameters.
Once the parameters $A,B$ and $M$ are given, the coefficients $a_i$ are found by minimizing the expectation value of the Hamiltonian, 
$E_{\Psi_{GEM}}=\langle \Psi_{GEM}|H|\Psi_{GEM}\rangle/\langle \Psi_{GEM}|\Psi_{GEM}\rangle$. We vary the parameters $A,B$ and $M$ 
to find the minimal value of $E_{\Psi_{GEM}}$, which is used in the main text as $E$.  
To benchmark our numerical calculations, we used known results for charged and neutral systems~\cite{varga1995, suzuki2002,blume2014}.
In addition, we cross-checked some of the energies presented in the main text using the stochastic variational method with correlated Gaussians (SVM)~\cite{suzuki1998, mitroy2013}.
In our implementation, the SVM assumes the variational wave function 
in the form 
\begin{equation}
\Psi_{SVM}=\sum_{i=1}^{M_S} a^S_i e^{-\alpha^S_i x^2-\beta^S_i y^2-\gamma^S_i {\bf x}\cdot{\bf y}},
\label{eq:SVM}
\end{equation}
where the parameters $\alpha^S_i$, $\beta^S_i$ and $\gamma^S_i$ [$\alpha^S_i\beta^S_i-\left(\gamma^S_i\right)^2/4>0$]
are found by a stochastic search (cf.~\cite{suzuki1998, mitroy2013}); $a^S_i$ are chosen 
to minimize the expectation value of the Hamiltonian. 
For weakly-bound neutral states one might restrain these parameters to better reproduce the tails 
of the wave function (cf.~\cite{volosniev2011a}). In our explorations, we saw that this option 
does not drastically improve convergence for charged systems, probably,
because the corresponding tails decay faster with distances than those of neutral systems.

Finally, we briefly explain why our results do not gain any systematic 
errors due to the choice of the basis. 
To this end, we show that an eigenstate with $L=0$ ($L$ for total angular momentum)
of the Hamiltonian can be accurately approximated by the variational wave function~(\ref{eq:SVM}).
This is not a trivial observation, since $\Psi_{SVM}$ depends only on the three variables: $x,y$, and $\theta_{xy}$ --
the angle between ${\bf x}$ and ${\bf y}$.  In general, an eigenstate might also depend 
on other combinations of angles that determine ${\bf x}$ and ${\bf y}$ 
(e.g., $\theta_x$,$\theta_y$, $\phi_x$, $\phi_y$ in spherical coordinates). 
A suitable angular basis for our discussion is $Y_{l_1 m_1}(\theta_y,\phi_y)Y_{l_2 m_2}(\theta_x,\phi_x)$,
where $Y$ is a real spherical harmonic. 
Any suitable variational function can be written as
\begin{equation}
\Phi_{m_1,m_2}=\sum_{i,l_1,l_2} f_{l_1,l_2}^{i}(x,y) Y_{l_1 m_1}(\theta_y,\phi_y)Y_{l_2 m_2}(\theta_x,\phi_x),
\end{equation}
where $f$ is the function that determines the expansion. We note that the Hamiltonian, $H_3$, does not mix the subset 
$\{P_l({\bf x}\cdot{\bf y})\}$ with the rest of the basis, 
$P_l$ is the Legendre polynomial [$P_l({\bf x}\cdot{\bf y})=\frac{4\pi}{2l+1}\sum_{m=-l}^{l}Y_{lm}(\theta_y,\phi_y)Y_{lm}(\theta_x,\phi_x)$]. 
Indeed, since two-body potentials $V$ depend only on $x,y$ and $\theta_{xy}$, we derive
\begin{equation}
H_3 P_l({\bf x}\cdot{\bf y}) =\sum_{l'} F_{l'}(x,y)P_{l'}({\bf x}\cdot{\bf y}),
\end{equation}
where $F_{l'}$ is an irrelevant for our discussion function that depends on $x$ and $y$. Therefore, there are eigenstates of $H_3$ 
that can be written as
\begin{equation}
\phi=\sum f_{l}^i(x,y) P_l({\bf x}\cdot{\bf y}). 
\end{equation}
It is expected that a square integrable function $\phi$ can be well represented by a suitable 
$\Psi_{SVM}$~\cite{suzuki1998}. This becomes intuitively clear after writing Gaussian functions 
in the form of the corresponding Maclaurin series. One can confirm that $\Psi_{SVM}$ describes the ground state 
by checking numerically that it does not change sign. For the considered cases, variational results using $\Psi_{SVM}$ and $\Psi_{GEM}$
[the Gaussian expansion method] agree well, which means that the form of $\Psi_{GEM}$ can approximate accurately the function $\phi$.  
The dependence on $\theta_{xy}$ in $\Psi_{GEM}$ appears due to the presence of the symmetrization operator $\mathcal{S}$.

\section{Thomas Collapse}
\label{app:thomas}

\begin{figure}
\centerline{\includegraphics[scale=0.6]{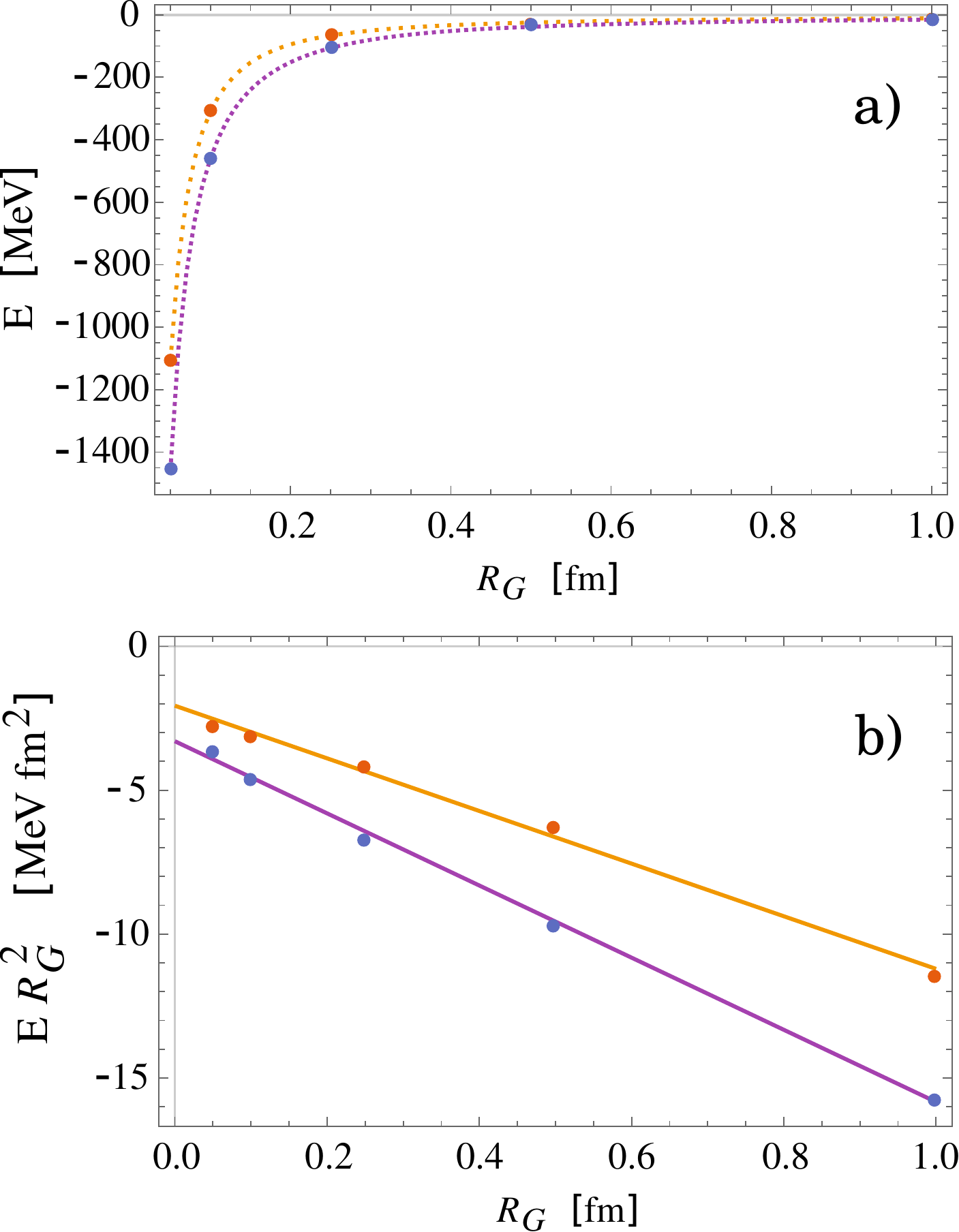}}
\caption{An illustration of the Thomas collapse for three charged particles. 
Panel {\bf a)}: The three-body energy as a function $R_G$
for charged particles (see lower dots);
the curved line is added to guide an
eye. For comparison, we also plot the three-body energies for the corresponding neutral system
(upper dots).
 Panel {\bf b)}: The three-body energy times $R_G^2$ as a function $R_G$
for charged particles (see lower dots); 
the line shows the corresponding linear fit. 
For comparison, we also plot results for the corresponding neutral system
(upper dots).}
\label{fig:thomas_collapse}
\end{figure}

If a zero-range two-body potential is set to reproduce a 
binding energy of two neutral particles, then the ground state of the corresponding three-body system is
infinitely deep. This phenomenon is called the Thomas collapse~\cite{thomas1935}. 
A way to deal with this peculiarity in models with zero-range potentials is to introduce a three-body parameter.
Here we show numerically that calculations with three charged particles, even in spite of a repulsive Coulomb potential, also require a three-body force. 
To this end, we compute three-body energies for two-body Gaussian potentials that have different values $R_G$ but lead to the same value of the two-body binding energy.
For the sake of discussion, we use masses and charges of $\alpha$-particles, and assume that the
 binding energy is $1$ MeV. The results are presented in Fig.~\ref{fig:thomas_collapse}.
The three-body bound state becomes unphysically deep for $R_G\to 0$ (the right
panel of the figure suggests that $E_3\sim 1/R_G^2$ for $R_G\to 0$), which shows the necessity of a three-body 
force.

\begin{figure*}
\centerline{\includegraphics[scale=1.1]{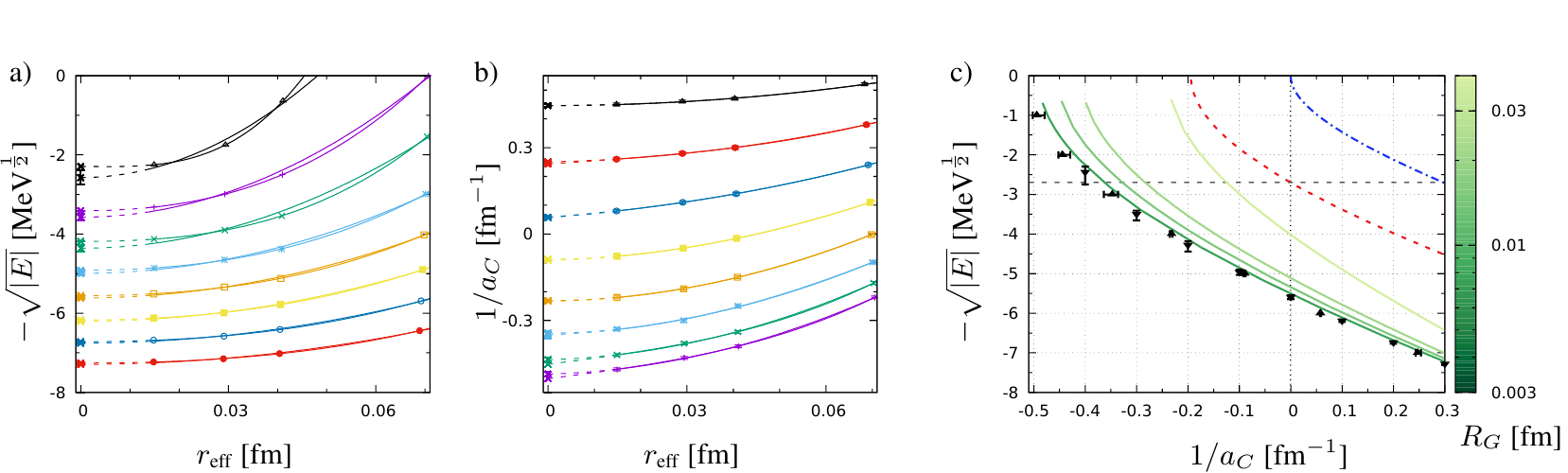}}
  \caption{An illustration of the extrapolation scheme employed to calculate the tetramer energies in Fig.~2 of the main text. 
 Panels~{\bf a)}~and~{\bf b)} show the extrapolation fits while panel {\bf c)} shows the 
 curves for finite values of $R_G$ and the result of the extrapolation to $R_G \to 0$.
 Panel~{\bf a)}: The points are taken from the curves in panel {\bf c)} at fixed values of $1/a_C$. From the top to bottom the 
 points and curves correspond to $1/a_C=-0.4, -0.3, ... 0.3\,\text{fm}^{-1}$. The fit functions are explained in the text. The dashed curves 
 show where the fit functions are used for extrapolation. The points with error bars at $r_\text{eff}=0$ represent the result of the
 extrapolation. Panel {\bf b)}: Same as panel {\bf a)}, but at a fixed energy $E_4$, $E_4=-8,-7,..-1\,\text{MeV}$ from the top to bottom. 
 Panel~{\bf c)}: The results for the tetramer binding energy vs. the inverse scattering length at different $R_G$. 
Smaller values of $R_G$ are represented with a darker green color. The red dashed line shows the zero-range result for the trimer and 
 the blue dot-dashed line shows the zero-range result for the dimer, see the main text for detail. 
The black dots with error bars correspond to the extrapolated points in panels {\bf a)} and {\bf b)}.}
 \label{fig6}
\end{figure*}

\section{Extrapolation Procedure}
\label{app:extra}

To obtain results in the limit $R_G\to 0$, we first calculate energies for a sequence of small
values of $R_G$ and then use the extrapolation procedure described below. 
For trimers, the finite-range corrections to the energy for the smallest numerically accessible values of $R_G$ 
are negligible. Therefore, we present the procedure only for tetramers. 
The tetramer energies for the four smallest $R_G$ ($R_G = 0.0075\,\text{fm}, 0.0150\,\text{fm},
0.0212\,\text{fm}$ and $0.0374\,\text{fm}$) are presented in 
Fig.~\ref{fig6}{\bf c)}. The three-body forces are taken from the corresponding trimer calculations. 
The figure shows that even a marginal change from $R_G=0.0075\,\text{fm}$ to $R_G=0.015\,\text{fm}$
changes noticeably the energies. To obtain energies for $R_G\rightarrow 0$ and error bars in the $x$ and $y$ direction,
we extract data from Fig.~\ref{fig6}{\bf c)} at fixed values of $1/a_C$ and at fixed values of $E$.
These data are shown in Figs.~\ref{fig6}{\bf a)}~and~{\bf b)} for the four smallest values of $R_G$. 
Then we employ the fit function
\begin{equation}
 f(r_\text{eff}) = a_1r_\text{eff} + c_1r_\text{eff}^2 + b_1
\end{equation}
which assumes an analytical functional dependence on $r_\text{eff}$ near $r_\text{eff}=0$. 
In addition, we use the fit function
\begin{equation}
 \tilde f(r_\text{eff}) = a_2 r_\text{eff} +  c_2 r_\text{eff}^2 + d_2 r_\text{eff}^3 + b_2,
\end{equation}
which includes an additional $r_\text{eff}^3$ term in comparison to $f$.
The fits with $f$ and $\tilde f$ yield $b_1$ and $b_2$, and the error bars on $b_1$ and $b_2$, which we refer to as $e_1$ and $e_2$.
Note that $e_2=0$ because we use only four points for the fit. The average between $b_1$ and $b_2$ is used as a central value for the dots in Fig.~\ref{fig6}{\bf c)} (see also Fig.~2 of the main text).
The error bars correspond to the smallest/largest value of $b_i-e_i$ / $b_i+e_i$, 
respectively. The result of this procedure is shown in Fig.~\ref{fig6}{\bf c)}, where each point 
corresponds either to a vertical or horizontal extrapolation. 
The figure shows that the two extrapolation directions agree well.


\end{document}